\documentclass[aps,prc,twocolumn,times,graphicx,tighten,showpacs]{revtex4}
\usepackage[dvips]{graphicx}
\usepackage[dvips]{color}
\usepackage{amsmath}

\newcommand{\bea}{\begin{eqnarray}}
\newcommand{\eea}{\end{eqnarray}}
\newcommand{\be}{\begin{equation}}
\newcommand{\ee}{\end{equation}}
\newcommand{\np}{{\bf p}}

\newcommand{\nh}{{\bf h}}

\newcommand{\tauvec}{\mbox{\boldmath $\tau$}}
\newcommand{\Ivec}{\mbox{\boldmath $I$}}

\newcommand{\sumint}{\sum\kern -3.5ex \int\kern 1.0ex}

\newlength\dlf  % Define a new measure, dlf

\begin{document}

%------------------------------------------
\title{
Emission of 
neutron-proton and proton-proton pairs in neutrino scattering 
}

%--------------------
\author{
I. Ruiz Simo$^a$,
J.E. Amaro$^a$,
M.B. Barbaro$^{b,c}$,
A. De Pace$^c$,
J.A. Caballero$^d$,
G.D. Megias$^d$, 
T.W. Donnelly$^e$
}
%--------------------

\affiliation{$^a$Departamento de F\'{\i}sica At\'omica, Molecular y Nuclear,
and Instituto de F\'{\i}sica Te\'orica y Computacional Carlos I,
Universidad de Granada, Granada 18071, Spain}

\affiliation{$^b$Dipartimento di Fisica, Universit\`a di Torino,
 Via P. Giuria 1, 10125 Torino, Italy}

\affiliation{$^c$INFN, Sezione di Torino, Via P. Giuria 1, 10125 Torino, Italy}  

\affiliation{$^d$Departamento de F\'{\i}sica At\'omica, Molecular y Nuclear,
Universidad de Sevilla, Apdo.1065, 41080 Sevilla, Spain}

\affiliation{$^e$Center for Theoretical Physics, Laboratory for Nuclear
  Science and Department of Physics, Massachusetts Institute of Technology,
  Cambridge, MA 02139, USA}

\date{\today}

%----------------------------------------------------------------------

\begin{abstract}

We use a recently developed model of relativistic meson-exchange
currents to compute the neutron-proton and proton-proton yields in
$(\nu_\mu,\mu^-)$ scattering from $^{12}$C in the 2p-2h channel. We
compute the response functions and cross sections with the relativistic
Fermi gas model for different kinematics from intermediate to high
momentum transfers.  We find a large contribution of neutron-proton
configurations in the initial state, as compared to proton-proton
pairs. In the case of charge-changing neutrino scattering the 2p-2h
cross section of proton-proton emission ({\it i.e.,} np in the initial
state) is much larger than for neutron-proton emission ({\it i.e.,} two neutrons
in the initial state) by a $(\omega,q)$-dependent factor.  The
different emission probabilities of distinct species of nucleon pairs are
produced in our model only by meson-exchange currents, mainly by the
$\Delta$ isobar current. We also analyze other effects including exchange
contributions and the effect of the axial and vector currents.

\end{abstract}

%-----------------------------------
\pacs{25.30.Fj; 21.60.Cs; 24.10.Jv}
%-------------------------------------

\maketitle

\section{Introduction}

The identification of nuclear effects in neutrino scattering is
essential for modern neutrino oscillation experiments
\cite{Agu10,Nak11,And12,Abe13,Fio13,Abe14,Wal15}.  In particular the
sensitivity of the neutrino energy reconstruction to multi-nucleon
events has been stressed in recent data analyses \cite{Rod16}.  In the
MINERvA neutrino experiment an enhanced population of multi-proton
states has been observed between the quasielastic and $\Delta$ peaks.
On the other hand, observation of events with a pair of energetic
protons at the interaction vertex accompanying the muon in
$^{40}$Ar$(\nu_\mu,\mu^-)$ reaction has been reported in the ArgoNeuT
experiment \cite{Acc14}. From these events several back-to-back
nucleon configurations have been identified and associated with nuclear
mechanisms involving short-range correlated (SRC) neutron-proton (np)
pairs in the nucleus \cite{Cav15}.  However in \cite{Wei16} these
``hammer events'' have been modeled by a simple pion production and
reabsorption model without { nucleon-nucleon} correlations, suggesting that the
distribution of pp pairs in the final state is less sensitive to
details of the initial pair configuration. In the opinion of the
authors of \cite{Wei16}, the events cannot teach us anything
significant about SRC. The NUWRO event generator supports that the
excess of back-to-back events in ArgoNeuT has a kinematic origin and
is not directly related to SRC \cite{Nie16}.

SRC with back-to-back configurations have been also identified in
two-nucleon knock-out electron scattering experiments on $^{12}$C for
high momentum transfer and missing momentum \cite{Shn07,Sub08}. In
this case one expects an excess of np pairs over pp pairs
\cite{Ryc15,Col16}.  The experiment reported a number of
np pairs 18 times larger than their pp counterparts.  The
analysis of these experiments is compatible 
with theoretical single-nucleon
and nucleon pair momentum distributions in variational Monte Carlo
calculations, where the importance of the tensor forces in the ground-state correlations of nuclei has been emphasized \cite{Sch07,Wir14}.
While the kinematics of the experiments have been selected to minimize
the contribution from other mechanisms that can induce two-particle
emission, such as meson-exchange currents (MEC) and isobar excitations
\cite{Shn07}, the contribution of MEC cannot be ruled out {\em a
  priori} \cite{Sim16b}.

In this work we investigate the relative effect of MEC on the separate
pp and np channels in the inclusive 2p-2h neutrino cross section.  It
has been emphasized that the separate charge distributions of 2p-2h
events are useful. One of the reasons is for their use in Monte Carlo
event generators \cite{And10,Nie16}.  For instance in NUWRO
configurations the MEC 2p-2h excitations are assumed to occur 95\% of
the time for events where the interaction occurs in initial np pairs
\cite{Nie16,Sob12} (or final pp pairs for charged current neutrino
scattering).  This value was estimated based on the assumption,
claimed also in \cite{Mar09}, that neutrinos interact mostly with
correlated np pairs.  From a naive calculation this value agrees with
a factor 18/19, corresponding to the extracted value of np/(np+pp) in
the $^{12}$C$(e,e'Np)$ experiment of \cite{Shn07}.  However this
neutrino generator uses a 2p-2h model that does not give separate pp
and np contributions, and therefore this choice is not fully
consistent from the theoretical point of view \cite{Nie16}. On the
other hand it is expected that the ratio between np and pp
interactions should be kinematics dependent and not only a global
factor. Thus a theoretical quantification of the np/pp ratio and its
dependence on the typical kinematics would be desired for each
implementation of 2p-2h cross sections.  Results for the separate pp
and np contributions due to short-range correlations have been
presented in \cite{Van16}, for the $R_T$ and $R_{CC}$ response
functions, and for $q=400$ MeV/c, but not for the differential cross
section.  The contribution of initial np pairs to the T response
found in \cite{Van16} is about twice that of the initial nn
pairs.

We have recently developed a fully relativistic model of
meson-exchange currents in the 2p-2h channel for electron and neutrino
scattering \cite{Sim16}. This model is an extension of the
relativistic MEC model of \cite{Pac03} to the weak sector. It has been
recently validated by comparing to the $^{12}$C$(e,e')$ inclusive
cross section data for a wide kinematic range within the SuperScaling
approach (SuSA) \cite{Meg16}. This model describes jointly the
quasielastic and inelastic regions using two scaling functions fitted
to reproduce the data, while the 2p-2h MEC contribution properly fills
the dip region in between, resulting in excellent global agreement
with the data.  The model has been recently extended to the
description of neutrino scattering reactions for a variety of
experiments providing an excellent agreement with data
\cite{SuSAv2-MEC-neutrino}.  With this benchmark model we are able to
study the separate np and pp channels in the response functions and
cross section for the three $(e,e')$, $(\nu_l,l^-)$ and
$(\bar{\nu}_l,l^+)$ reactions.  While this analysis was performed in
\cite{Sim16b} for electron scattering, in this work we consider
neutrino reactions. Our model includes the contributions of
pion-in-flight, seagull, pion-pole and $\Delta$(1232) excitation
diagrams of the MEC. The two-body matrix elements between relativistic
spinors were presented in our recent work \cite{Sim16}, where they
have been deduced from the weak pion production amplitudes of
\cite{Her07}.

\section{Formalism for neutrino scattering}

The formalism of 2p-2h cross section including MEC in the relativistic Fermi gas
was given in \cite{Sim16}. We write the charged current (CC))  
cross section as
\begin{eqnarray}
\frac{d\sigma}{d\Omega'd\epsilon'}
&=& 
\sigma_0 
\left[
\widetilde{V}_{CC} R^{CC}
+ 2 \widetilde{V}_{CL} R^{CL}
+ \widetilde{V}_{LL} R^{LL}
\right.
\nonumber\\
&&
\left.
+ \widetilde{V}_{T} R^{T}
\pm 2 \widetilde{V}_{T'} R^{T'}
\right],
\label{sigma}
\end{eqnarray}
where $\sigma_0$ is a kinematic factor including the weak couplings
defined in \cite{Ama05,Ama16}.  Note that there is a linear
combination of five response functions, labeled as $CC, CL, LL, T $
and $T'$.  The $T'$ response function contributes differently for
neutrinos (plus sign) than for antineutrinos (minus sign).
 The $\widetilde{V}_K$ factors are kinematic functions that were 
defined in \cite{Ama05,Ama16}.

The response functions $R^K(\omega,q)$ depend on the energy and momentum transfer. They are computed here in a relativistic Fermi gas (RFG)
model, with Fermi momentum $k_F$,  where they can be expanded as
the sum of one-particle one-hole (1p-1h), two-particle two-hole
(2p-2h), plus additional channels.
Here we are interested in
the 2p-2h channel, where two nucleons 
with momenta $\np'_1$ and $\np'_2$
are ejected out of the  Fermi sea, $p'_i>k_F$,
leaving two hole states in the daughter nucleus,
 with momenta $\nh_1$ and $\nh_2$  (with $h_i<k_F$).

The  2p-2h response functions are computed as
\begin{eqnarray}
R^{K}_{2p-2h}
&=&
\frac{V}{(2\pi)^9}\int
d^3p'_1
d^3h_1
d^3h_2
\frac{m^4_N}{E_1E_2E'_1E'_2}
\nonumber \\ 
&&
r^{K}(\np'_1,\np'_2,\nh_1,\nh_2)
\delta(E'_1+E'_2-E_1-E_2-\omega)
\nonumber\\
&&\times
\theta(p'_2-k_F)
\theta(p'_1-k_F)
\nonumber\\
&&\times 
\theta(k_F-h_1)
\theta(k_F-h_2),
\label{hadronic}
\end{eqnarray}
where the momentum of the second nucleon is fixed by  momentum conservation
inside the integral sign, $\bf p'_2= h_1+h_2+q-p'_1$,
$V$ is the volume of the system, 
$m_N$ is the nucleon mass, while $E_i$ and $E'_i$ are the 
energies of the holes
and particles, respectively.

Using energy conservation, the calculation of the inclusive 2p-2h
responses of Eq.~(\ref{hadronic}) for given energy and momentum transfer
$(\omega,q)$, is reduced to a seven-dimensional integral that is
computed numerically following the methods developed in \cite{Ruiz14,Ruiz14b}.
The main ingredient of the calculation is the set of five response
functions $r^{K}(\np'_1,\np'_2,\nh_1,\nh_2)$, for the elementary 2p-2h
transition.  These elementary response functions are written in terms
of the two-body MEC antisymmetrized matrix elements, summed over spin.
We separate the contributions of the different charge channels to the
response functions.  These can be $(np, pp)$ for neutrinos, and
$(np,nn)$ for antineutrinos.  In \cite{Sim16} we derived general
formulae for the separate np and pp response functions. 

The total CC MEC for neutrino scattering can be written as
\begin{eqnarray}
 j^\mu_{\rm MEC}
&=&\tau_{+}(1) \,J^\mu_1(1^\prime\,2^\prime;1\,2)+
\tau_{+}(2)\, J^\mu_2(1^\prime\,2^\prime;1\,2)\nonumber\\
&+& \left(I_V\right)_{+}\,J^\mu_3(1^\prime\,2^\prime;1\,2) ,
\label{MEC-isospin}
\end{eqnarray}
where $\tau_+=\tau_x+i\tau_y$ and we have defined the isospin operators
\begin{equation}
\left(I_V\right)_{\pm}= (I_V)_x\pm i (I_V)_y
\end{equation}
that stands for the $\pm$-component of the
two-body isovector operator
\begin{equation}
\Ivec_V = 
i \left[\tauvec(1) \times\tauvec(2)\right] .
\label{isospin_seagull}
\end{equation}
The isospin-independent two-body currents $J_1^\mu$, $J_2^\mu$, and
$J_3^\mu$, follow from the amplitudes of weak pion production model of \cite{Her07} and are written in  \cite{Sim16}.

\begin{figure}[tph]
\begin{center}
\includegraphics[scale=0.87,bb=200 460 420 720]{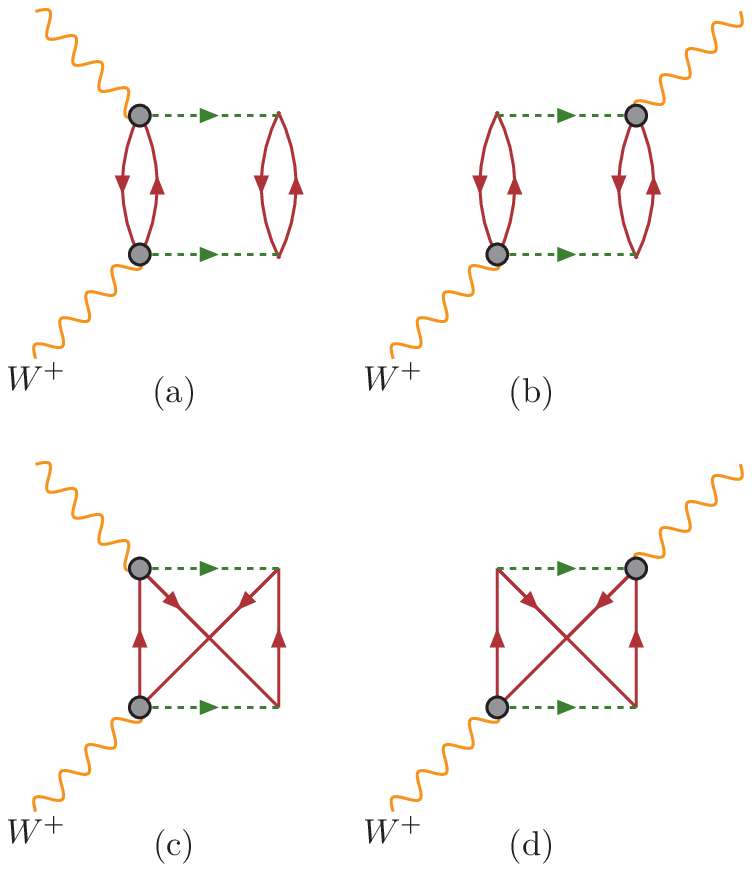}
\caption{(Color online) Some contributions of 2p-2h states to the
  response functions considered in this work.  The circle stands for
  the elementary $W^+ N \rightarrow \pi N$ amplitude. Diagrams (a,b)
  represent the direct contribution. Diagrams (c,d) are the exchange
  contributions.  }
\label{selfenergy}
\end{center}
\end{figure}

In other
models of neutrino scattering~\cite{Nie11,Mar09}, only
the direct diagrams (a,b) of Fig. \ref{selfenergy} are included,
while the direct-exchange contribution corresponding to the diagrams
(c,d) are disregarded.
In our model, on the contrary, both contributions are considered. 
The elementary 2p-2h transverse response function is given for pp emission by
\begin{eqnarray}
r^{T}_{pp}
&=& 
4\sum_{\mu=1}^2
 \sum_{s_1s_2s'_1s'_2}
\left\{ 
\left| 
J^\mu_{pp}(1'2';12)
\right|^2
\right.
\nonumber\\
&&
-{\rm Re}\;
J^\mu_{pp}(1'2';12)^*
J^\mu_{pp}(2'1';12)
\Big\} ,
\label{wPP}
\end{eqnarray} 
where $J^\mu_{pp}(1'2';12)$ is the 
effective two-body current for pp emission with neutrinos
given by 
\begin{equation}
J_{pp}^\mu = J_1^\mu+J_3^\mu. 
\end{equation}
The first term in the transverse response 
is the ``direct'' contribution, and the
second one is the ``exchange'' contribution, actually being the
interference between the direct and exchange matrix elements. 

The elementary transverse response for np emission has a similar expression 
\begin{eqnarray}
r^{T}_{np}
&=& 
4 
\sum_{\mu=1}^2
\sum_{s_1s_2s'_1s'_2}
\left\{ 
\left| 
J^\mu_{np}(1'2';12)
\right|^2
\right.
\nonumber\\
&&
-{\rm Re}\;
J^\mu_{np}(1'2';12)^*
J^\mu_{np}(1'2';21)
\Big\},
\label{wNP}
\end{eqnarray} 
but with the effective current $J^\mu_{np}$ for np emission
\begin{equation}
J_{np}^\mu = J_2^\mu+J_3^\mu ,
\end{equation}
 which is in general different from the pp one because of the distinct
 isospin matrix elements. The direct contribution corresponds to
 neglecting the second terms in Eqs. (\ref{wPP},\ref{wNP}).

\begin{figure*}
\begin{center}
\includegraphics[scale=0.77,bb=40 200 520 790]{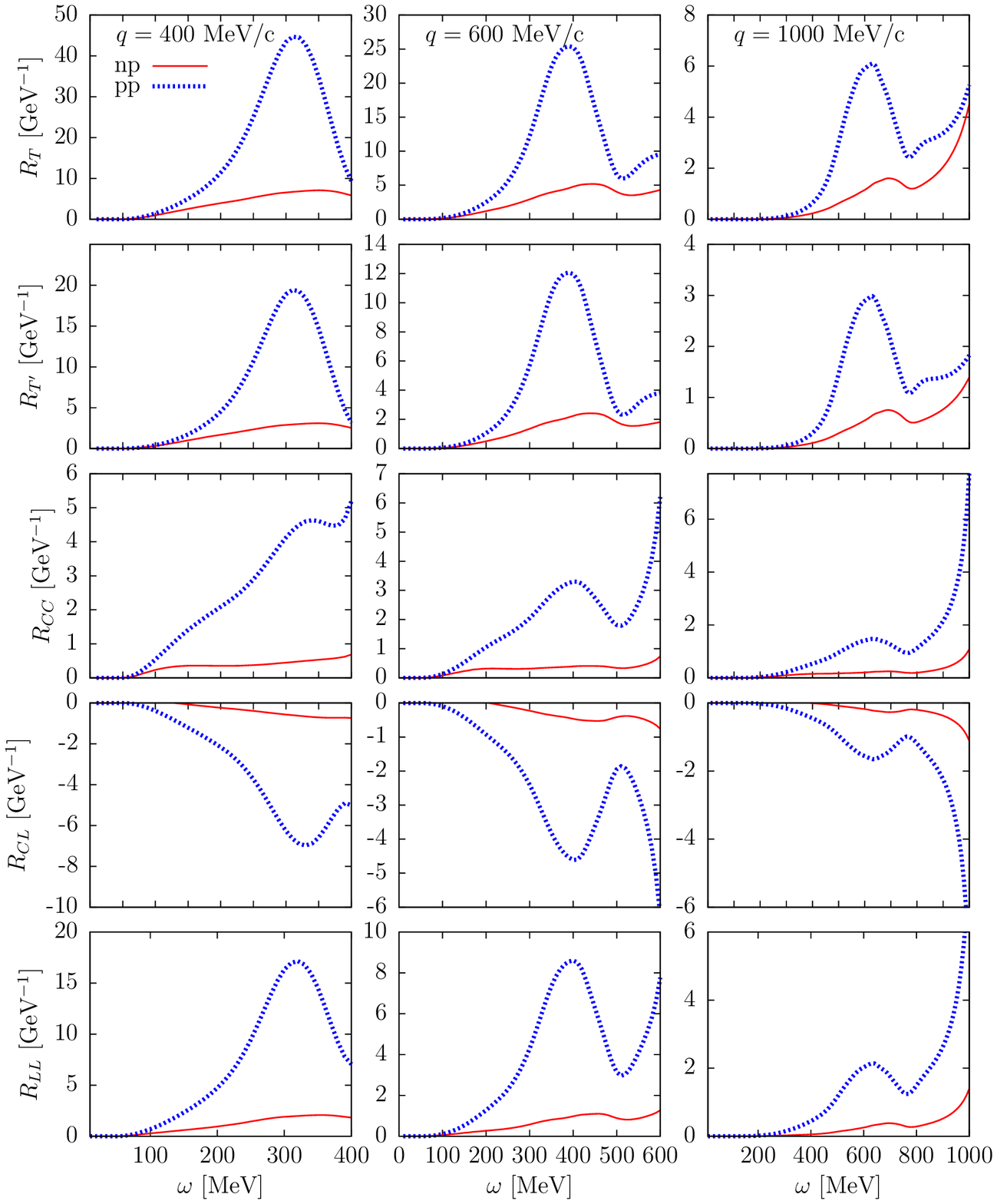}
\caption{(Color online) Separate pp and np 2p-2h response functions of
  $^{12}$C for three values of the momentum transfer. In all the
  figures the charge labels refer to the final pair of nucleons.  }
\label{fig1}
\end{center}
\end{figure*}

\section{results}

In the following we present results for the semi-inclusive
$^{12}$C$(\nu_\mu,\mu^-pp)$ and $^{12}$C$(\nu_\mu,\mu^-np)$ reactions,
corresponding to the 2p-2h channel of the $^{12}$C$(\nu_\mu,\mu^-)$
reaction in the two separate charge channels in the final hadronic
state corresponding to two-nucleon knockout. Our MEC model and its
parameters were obtained from the pion production amplitudes of \cite{Her07}. This is an extension of the electromagnetic MEC
model of \cite{Pac03} to the weak sector.

The five MEC-induced 2p-2h responses for CC neutrino interactions at
fixed momentum transfer are shown in Fig. \ref{fig1}. The Coulomb
$R_{CC}$ and transverse $R_T$ responses are also present in electron
scattering. In general the $\omega$ dependence of the 2p-2h responses
shows a broad peak coming from the $\Delta$ excitation. The strength
of the MEC peak weakens with $q$ due to the decrease of the
electroweak form factor with $Q^2$, especially the $\Delta$ form
factors. 
Note also that the most important contribution to the neutrino cross section
comes form the two transverse responses $R_T$ and $R_{T'}$.  
In the figure we only show the separate pp and np 2p-2h response functions,
the total responses being the sum of the two. 
For all the cases in Fig. \ref{fig1} we observe that 
the pp response functions are much larger than
  the np ones by a factor 6 or less depending on the kinematics.

The pp/np ratio in the present neutrino calculation 
can be compared to the np/pp ratio in the $(e,e')$
reaction studied in \cite{Sim16b}, because they correspond to the
same pairs in the initial state. For the transverse response 
that ratio for neutrino scattering is roughly a factor of two
smaller than for the electron case.

For $q=400$ MeV/c our results for 
$R_T$ can be compared to those of the
SRC model of \cite{Van16}.  Our MEC response at the maximum is
one order of magnitude larger than that of the SRC one.  In our
calculation, the pp pair emission transverse response induced by MEC
is about a factor of 6 larger than the np one. In contrast, the mentioned SRC
model shows at most a factor of 3 between 
 the two
contributions.  The
order of magnitude of the $R_{CC}$, on the other hand, is small in
both MEC and SRC 2p-2h responses, but still the MEC results are about
twice those of \cite{Van16}. The pp pairs in the final state
continue to dominate the $R_{CC}$ MEC response, while in the SRC case, both
pairs contribute similarly.

\begin{figure}
\begin{center}
\includegraphics[scale=0.95,bb=200 470 420 790]{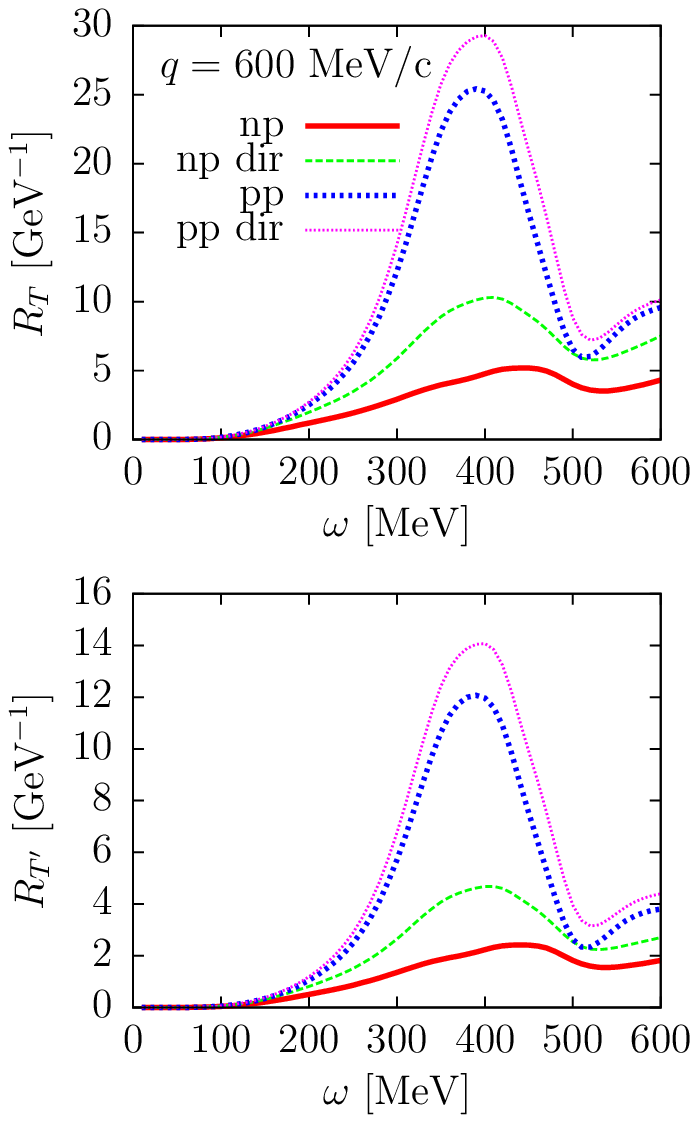}
\caption{(Color online) Separate    pp and np 
contributions to the $T$ and $T'$ 2p-2h response functions of
  $^{12}$C, for $q=600$ MeV/c, compared to the direct 
  contributions obtained by neglecting the direct-exchange interferences.  }
\label{fig3}
\end{center}
\end{figure}

Although the behaviors of MEC and SRC 2p-2h responses are completely
different, one should emphasize that the cross section is dominated by
the transverse responses. One could conclude from this comparison that
the MEC are the largest contribution to the 2N knockout strength, and
that the influence of SRC  is around 10\%.

 In a previous work \cite{Sim16} we showed that the interference
 diagrams (c,d) of Fig. 1 can amount to $\sim 25\%$ of the total 2p-2h
 responses.  But for the separate charge channels the interference
 influence can be truly different, as we show in Fig. \ref{fig3}.
 While the interference for pp emission produces a reduction of 20\%,
 for np emission the reduction factor is about 1/2.  Thus the ratio
 pp/np critically depends on the treatment on the interference
 contributions. The same conclusion was found for the electron
 scattering responses in \cite{Sim16b}.  The effect from the
 interference contribution is of the same size for the $T'$ response,
 as shown in the lower panel of Fig. \ref{fig3}.

\begin{figure}
\begin{center}
\includegraphics[scale=0.99,bb=200 320 420 790]{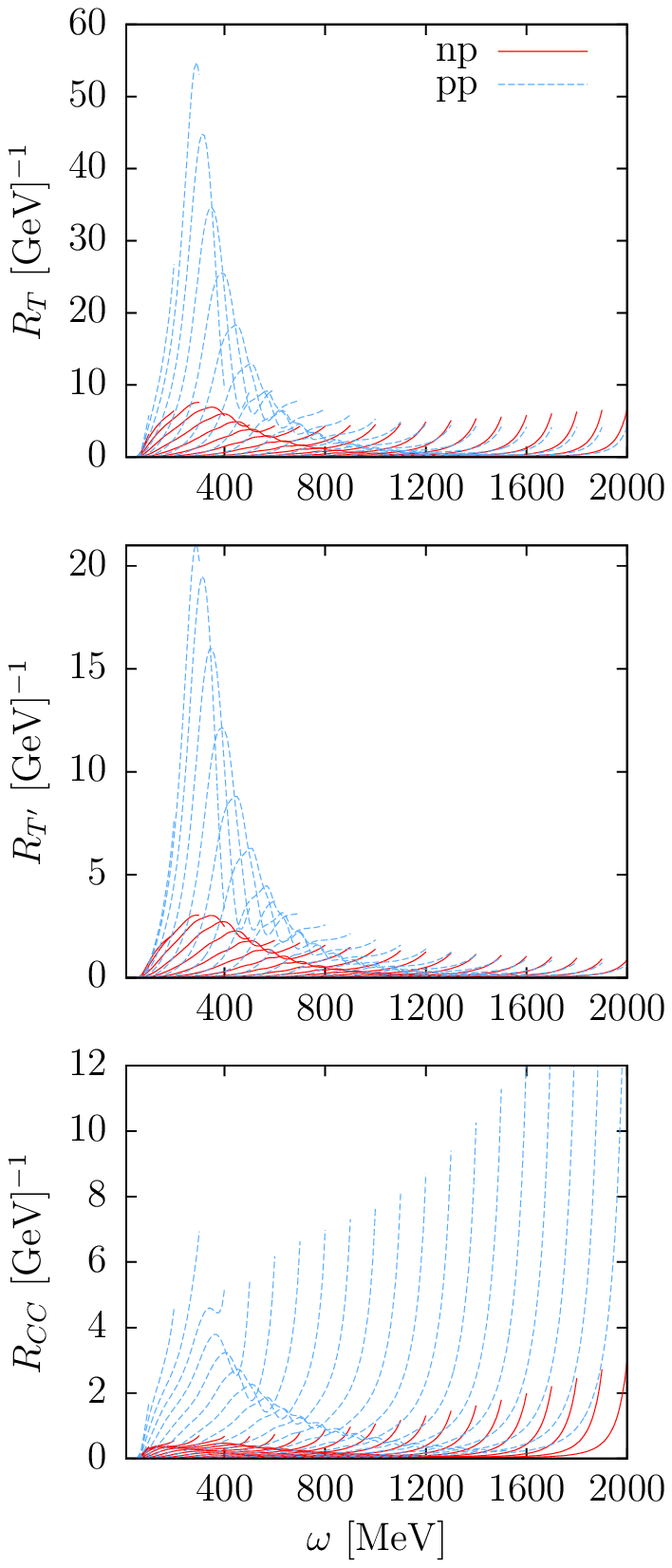}
\caption{(Color online)  
Comparison of the separate   pp and np contributions to the 
$T$, $T'$ and $CC$ 2p-2h
  response functions of $^{12}$C, for $q=100$, 200, ..., 2000 MeV/c.
}
\label{fig4}
\end{center}
\end{figure}

\begin{figure}
\begin{center}
\includegraphics[scale=0.99,bb=200 630 420 800]{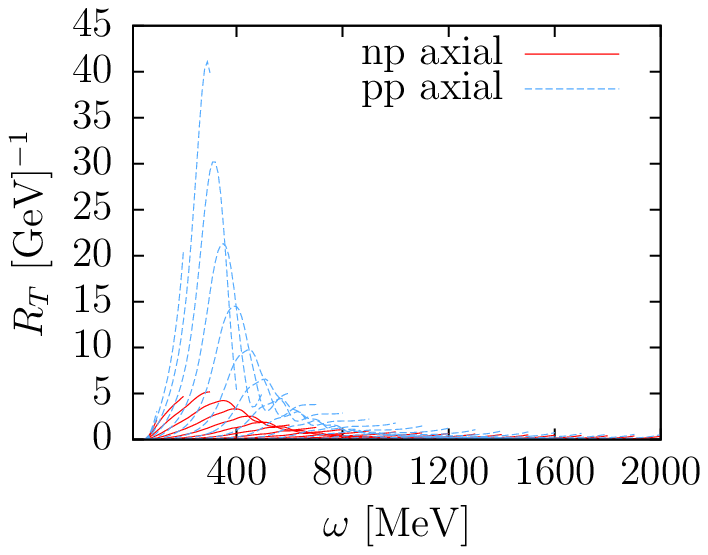}
\caption{(Color online)  
Comparison of the separate   pp and np contributions to the 
$T$  2p-2h
  response function of $^{12}$C, for $q=100$, 200, ..., 2000 MeV/c,
including only the axial MEC.
}
\label{fig5}
\end{center}
\end{figure}

In Fig. \ref{fig4} we compare the separate np and pp contributions to
the response functions $R_T$, $R_{T'}$, $R_{CC}$, in the full range of
momentum transfer from $q=100$ to 2000 MeV/c.  This corresponds to the
typical kinematic range of the neutrino experiments operating in the
few-GeV region. The $T$ response is more than twice the $T'$ both for
np and for pp channels. This seems to indicate that the axial MEC
contribution is larger than the vector one.  Indeed, this can be
truly observed in Fig. \ref{fig5}, where only the axial MEC current
is included in the calculation. Note that the $CC$ response is much
smaller than the other two, except for the $q=\omega$ point, where the
$C$ and $L$ contributions are approximately cancelled. They would be
exactly cancelled if the total current was conserved.  As a matter of
fact, near the photon point the seagull current dominates the MEC for
high $q$ when the $\Delta$ resonance is far away. The axial seagull
contribution is mainly longitudinal. That is why the $CC$ response in
Fig. \ref{fig4} is large at the photon point for high $q$ and the $T$
response in Fig. \ref{fig5} is so small.

\begin{figure}
\begin{center}
\includegraphics[scale=0.99,bb=200 240 420 780]{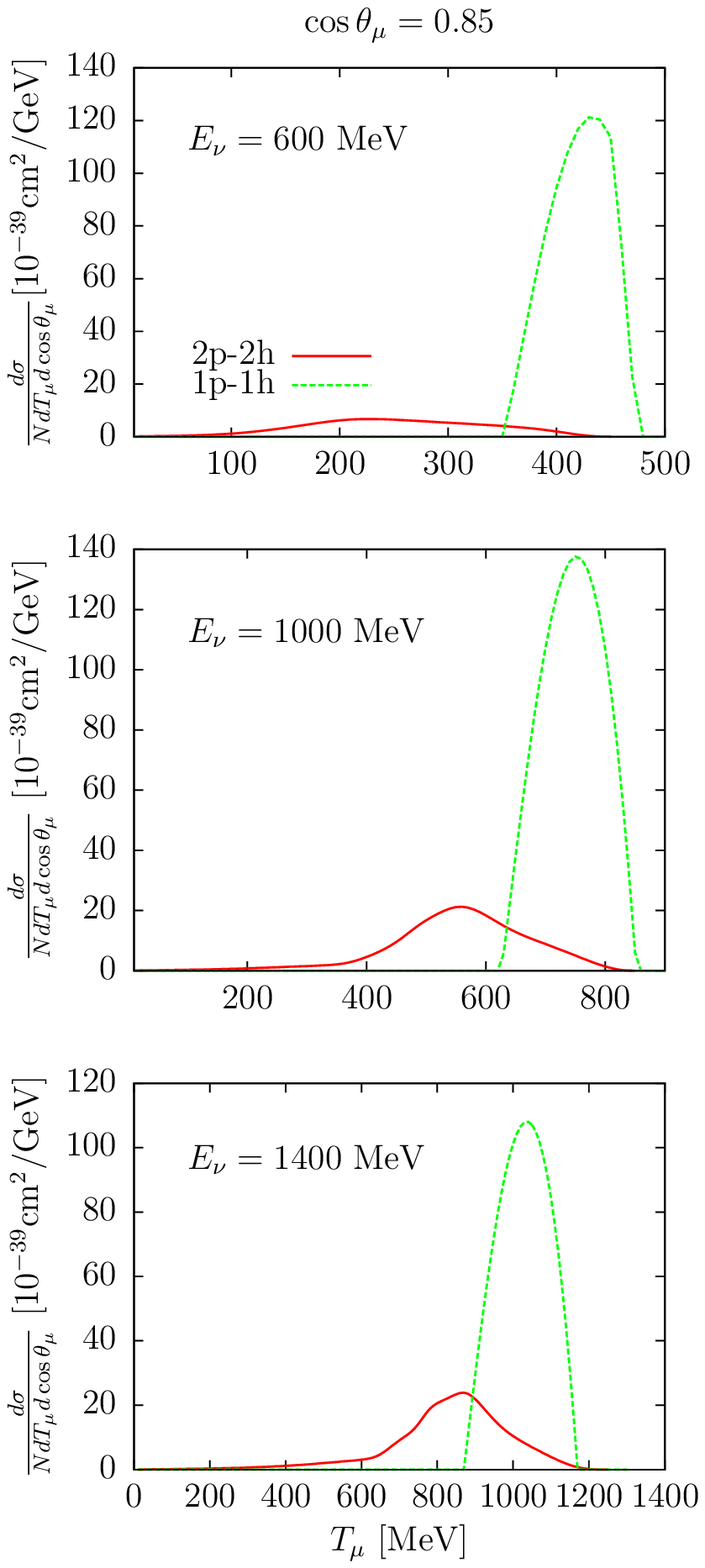}
\caption{(Color online) Double differential neutrino cross section per
  neutron of $^{12}$C, for fixed muon scattering angle and for three
  neutrino energies, as a function of the muon kinetic energy.  The
  separate 1p-1h and 2p-2h cross sections are displayed for
  comparison.  }
\label{fig6}
\end{center}
\end{figure}

\begin{figure}
\begin{center}
\includegraphics[scale=0.99,bb=200 240 420 780]{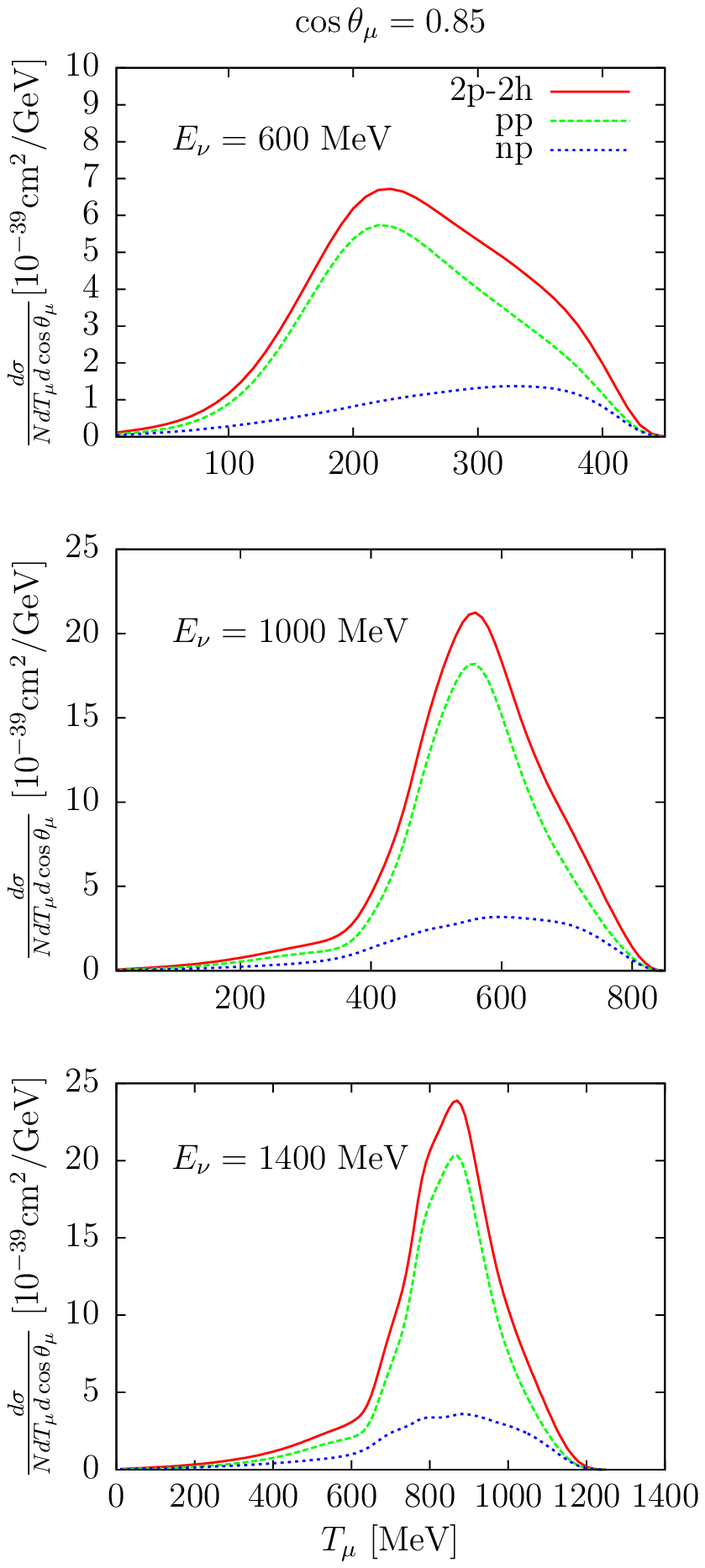}
\caption{(Color online) Double differential 2p-2h neutrino cross section per neutron of
  $^{12}$C, for fixed muon scattering angle and for three neutrino
  energies, as a function of the muon kinetic energy.  The separate
  np and pp channels are shown.  }
\label{fig7}
\end{center}
\end{figure}

To appreciate the size of the MEC 2p-2h contribution, in
Fig. \ref{fig6} we plot the double differential neutrino cross section
 per neutron, $d^2\sigma/d\cos\theta_\mu / dT_\mu/N$, of $^{12}$C, as a
function of the muon kinetic energy, for $\cos\theta_\mu=0.85$ and for
three values of the incident neutrino energy. We show the separate
contributions of 1p-1h and 2p-2h channels in the RFG. The relative
contribution of 2p-2h increases with the neutrino energy, and the MEC
and quasielastic peaks get closer.  Here the neutrino energy is fixed,
while in the experiments the neutrino energy is not fixed and the flux
produces an average of all the contributions around the mean
energy. For typical peak energies of 1 GeV, the results of
Fig. \ref{fig6} indicate that one can expect a contribution of
roughly 20\% to the cross section from 2p-2h.

\begin{figure}
\begin{center}
\includegraphics[scale=0.99]{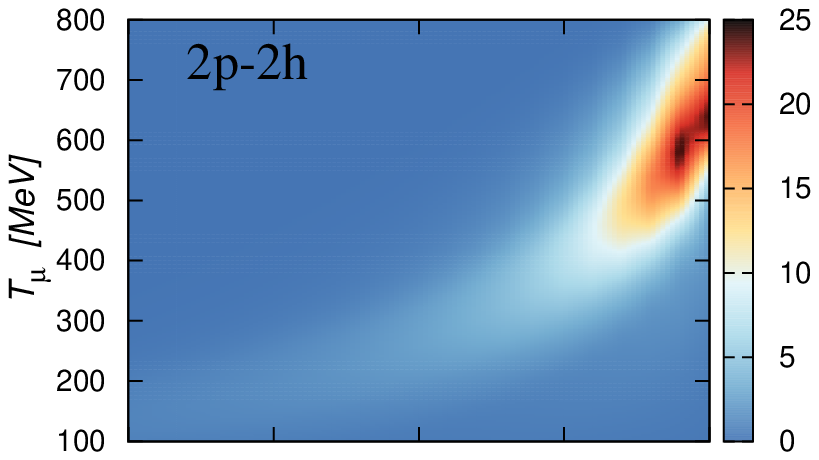}\\[-15mm]
\includegraphics[scale=0.99]{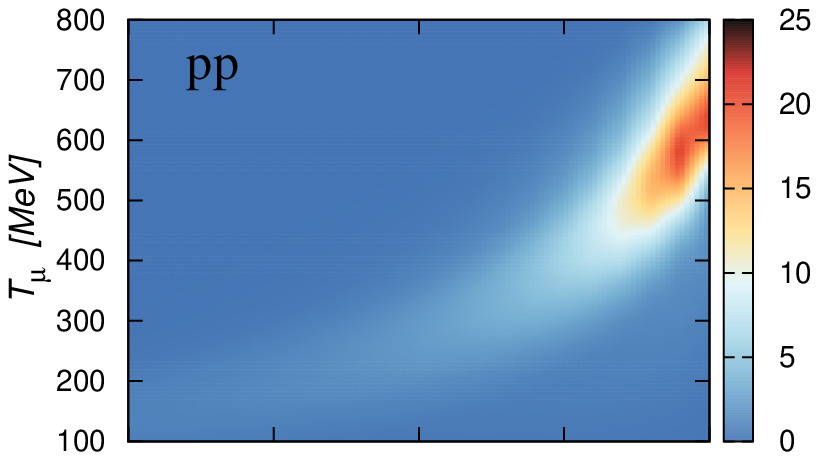}\\[-15mm]
\includegraphics[scale=0.99]{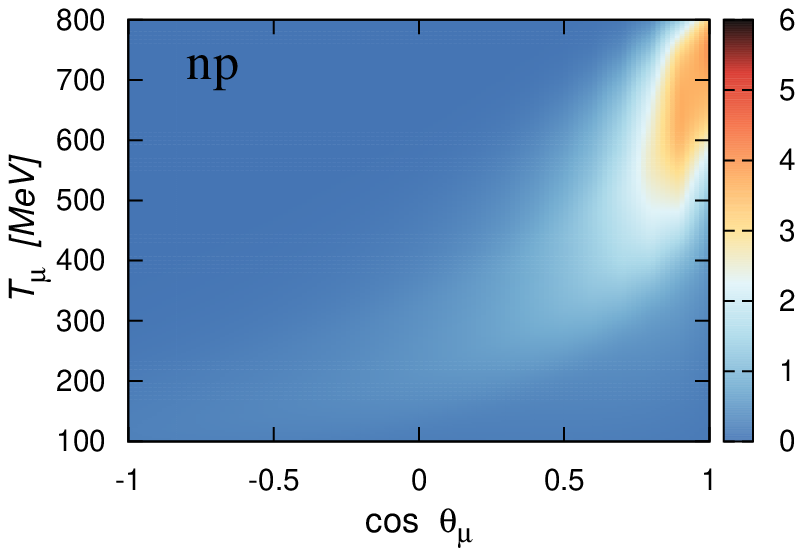}
\caption{(Color online)  
Double differential 2p-2h neutrino cross section per neutron of
  $^{12}$C, $d^2\sigma/d\cos\theta_\mu dT_\mu/N$, in units of 
$10^{-39}\rm cm^2/GeV$, as a function of 
$\cos\theta_\mu,T_\mu$ for fixed neutrino energy $E_\nu=1$ GeV.
 The separate   
 pp and np channels are shown in the middle and bottom panels, respectively. }
\label{fig8}
\end{center}
\end{figure}

 The separate pp and np channels in the differential neutrino cross
 section are shown in Fig. \ref{fig7} for the same kinematics. The pp
 channel clearly dominates the 2p-2h cross section. The pp/np ratio is
 around 5-6 near the maximum, but its precise value depends on the
 kinematics. Note that the np distribution is shifted towards higher
 muon energies compared with the pp case. 
This effect can be further
 observed in Fig. \ref{fig8}, where we show the
 $(\cos\theta_\mu,T_\mu)$ dependence of the 2p-2h double differential
 cross section, for $E_\nu=1$ GeV. Indeed the second and third panels
 show the separate pp and np distributions. The np is much smaller
 than the pp one, and it is clearly shifted towards higher $T_\mu$ and
 smaller angles. It can be seen that for this neutrino energy, the
 absolute maximum of the cross section is located around
 $\cos\theta_\mu \sim$ 0.85 and $T_\mu \sim 600$ MeV, and corresponds
 approximately to the maximum shown in the middle panel of
 Fig. \ref{fig7}. The 2p-2h strength is concentrated in the top-right
 corner of Fig. \ref{fig8} corresponding to small angles and large
 muon kinetic energies, meaning low energy transfer, around
 $\omega=300$ MeV. This corresponds to the excitation energy of the
 $\Delta(1232)$, which gives the main contribution to the MEC.  Our
 calculation predicts that, when the lepton scattering angle
 increases, two particle emission implies a decrease of the
 kinetic energy of the muon or larger $\omega$.

\section{Conclusions}

In this work we have studied the separate charge channels
$(\nu_\mu,\mu^-pp)$ and $(\nu_\mu,\mu^-np)$, from $^{12}$C, integrated
over the two emitted nucleons, that contribute to the 2p-2h cross
section in  
quasielastic-like CC neutrino scattering.  We have computed the
response functions and double differential cross sections for several
kinematics. The pp channel dominates over the np contribution in the
whole domain. The pp/np ratio is about 5-6 for a wide range of
 neutrino energies. Future plans are to fold the cross 
 section with the neutrino fluxes
for the
various neutrino oscillation experiments.
Having the separate isospin
contributions will allow us to apply this formalism to asymmetric
nuclei $N\ne Z$. This will be of interest for neutrino
experiments based, for instance, on $^{40}$Ar, $^{56}$Fe or
$^{208}$Pb.

%%%%%%%%%%%%%%%%%%%%%%%%%%%%%%%%%%%%%%%%%%%%%%%%%%%%%%%%%%%%%%%%%%%%%
\section*{Acknowledgments}
This work was supported by Spanish Direccion General de Investigacion
Cientifica y Tecnica and FEDER funds (grants No. FIS2014-59386-P and
No. FIS2014-53448-C2-1), by the Agencia de Innovacion y Desarrollo de
Andalucia (grants No. FQM225, FQM160), by INFN under project MANYBODY,
and part (TWD) by U.S. Department of Energy under cooperative
agreement DE-FC02-94ER40818. IRS acknowledges support from a Juan de
la Cierva fellowship from Spanish MINECO. GDM acknowledges support from a Junta de Andalucia fellowship (FQM7632, Proyectos de Excelencia 2011).
%%%%%%%%%%%%%%%%%%%%%%%%%%%%%%%%%%%%%%%%%%%%%%%%%%%%%%%%%%%%%%%%%%%%

\end{document}